\documentclass[a4paper,10pt,twocolumn,oneside]{article}

\usepackage[paper=a4paper,dvips,top=2.5cm,left=2.5cm,right=2.5cm,bottom=2.5cm]{geometry}

\usepackage[latin1]{inputenc}
\usepackage[T1]{fontenc}

\usepackage{epsfig}
\usepackage{amsmath}
\usepackage{amsfonts}
\usepackage{amssymb}

\usepackage{graphicx}

\begin{document}

\pagestyle{empty}

\newtheorem{example}{Example}
\newtheorem{definition}{Definition}
\newtheorem{theorem}{Theorem}
\newtheorem{corollary}{Corollary}
\newtheorem{lemma}{Lemma}
\newtheorem{Construction}{Construction}
\newtheorem{conjecture}{Conjecture}

\newcommand{\convolc}[1]{\bigcirc \!\!\!\!\! \textrm{{\small #1}} \,}
\newcommand{\conveqn}[1]{\bigcirc \!\!\!\!\!\! \textrm{{\small #1}} \ }
\newcommand{\convtbl}[1]{\bigcirc \!\!\!\!\!\!\!\: \textrm{{\tiny #1}} \: \,}
\newcommand{\define}{\stackrel{\Delta}{=}}

\title{\textbf{New Algorithms for Computing a Single Component of the Discrete Fourier Transform}}

\author{{G. Jer\^{o}nimo da Silva Jr., R. M. Campello de Souza and H. M. de Oliveira}
\\ Dept. of Electronics \& Systems\\UFPE, CP7800, 50711-970 Recife PE, Brasil
\\ E-mail: gilsonjr@gmail.com.\\}

\date{}


\maketitle

\begin{abstract}
This paper introduces the theory and hardware implementation of two new algorithms for computing a single component of the discrete Fourier transform. In terms of multiplicative complexity, both algorithms are more efficient, in general, than the well known Goertzel Algorithm. 
\end{abstract}


\thispagestyle{empty}

\section{Introduction}


Discrete transforms are mathematical tools used in many applications in Engineering. A particularly significant example is the discrete Fourier transform (DFT) \cite{OppenPDS}.
Let $v=(v_n)$, $n=0,\ldots, N-1$, be a sequence of complex numbers or of real numbers. The DFT of $v$ is the sequence of complex numbers $V=(V_k)$, $k=0,\ldots, N-1$, defined by
\begin{equation}\label{eq:DFT}
V_k\define \sum_{n=0}^{N-1}v_n W_N^{k n},
\end{equation}
where $W_N=e^{-j\frac{2\pi}{N}}$ and $j=\sqrt{-1}$.

 
The polynomial representation for an input signal $v$, denoted by $v(x)$, is defined by
\begin{equation}
v(x)\define \sum_{n=0}^{N-1}v_n x^{n}.
\end{equation}
Therefore, the component $V_k$ can be computed from $v(x)$ by
\begin{equation}\label{eq:Vk_pol}
V_k = v(W_N^k).
\end{equation}

From \eqref{eq:DFT}, the computation of a single coefficient $V_k \in \mathbb{C}$, requires $N-1$ complex multiplications, $N-1$ complex additions and the prestorage of the coefficients $W_N^{n k}$. An algorithm to implement this computation, without the need for storing the coefficients, was  presented in \cite{Goertzel}. The Goertzel algorithm, as it became known, computes the component $V_k$ via the polynomial
\begin{eqnarray}
p_k(x) = (x-W_N^k)(x-W_N^{-k})\nonumber\\
= 1-2\cos \left( \frac{2\pi k}{N} \right)x+x^2,
\end{eqnarray}
which is the minimal polynomial of $W_N^k$ over the field of real numbers. It is possible to write $v(x)$ as
\begin{equation}
v(x) = p_k(x)q(x)+r(x),
\end{equation}
where $q(x)$ and $r(x)$ are obtained by polynomial division. Since $p_k(x)$ has a zero in $W_N^k$,  \eqref{eq:Vk_pol} can be used to derive
\begin{equation}\label{eq:Vk_r}
V_k = r(W_N^k).
\end{equation}

\begin{figure}[hbt]
 \centering
 \includegraphics[scale=0.6]{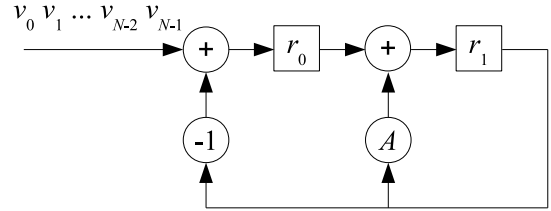}
 \caption{Autoregressive filter to compute the polynomial division by $p_k(x)$. The component $v_{N-1}$ is the	first to be fed into the shift register circuit and $A = 2\cos(2\pi k/N)$. The DFT component is computed by $V_k = r_0 + r_1 W_N^k$.}
 \label{fig:goertzel}
\end{figure}

If $v$ has real coefficients, the polynomial division by $p_k(x)$ requires $N-2$ real multiplications. Two real multiplications are necessary to compute $r(W_N^k)$, so that the Goertzel algorithm requires $N$ real multiplications to compute one component of an $N$-point DFT. The polynomial division can be implemented by an autoregressive filter, as shown in Figure \ref{fig:goertzel}.

Although the Goertzel algorithm can be used to compute the DFT of a given sequence, it is not a fast Fourier transform because its computational complexity, for an $N$-point DFT, is proportional to $N^2$. It is an attractive procedure for application scenarios where only a few components (not more than $\log_2 N$ of the $N$ components) of the DFT need to be computed, such as in the detection of DTMF signals \cite{Beck_Goertzel}. 

Cyclotomic polynomials play an important role in the algorithms introduced in this paper. The $N$-th cyclotomic polynomial, denoted by $\Phi_N(x)$, is the monic polynomial which has as its roots all order $N$ elements in $\mathbb{C}$. Therefore
\begin{equation}\label{eq:PHI}
\Phi_N(x)\define\prod_{\theta |\textrm{ord}(\theta)=N}(x - \theta).
\end{equation}
and
\begin{equation}
\prod_{d|N}\Phi_d(x)=(x^N - 1).
\end{equation}

From the M\"{o}bius inversion formula \cite{McElieceFF}, it can be shown that
\begin{equation}
\Phi_N(x)=\prod_{d|N}(x^d - 1)^{\mu(N/d)},
\end{equation}
where $\mu(n)$ is the M\"{o}bius function \cite{Burton}

\begin{equation}
\mu(n)\define\left\{\begin{array}{ll} 1, & \textrm{if $n=1$;}\\
0, & \textrm{if $\exists e_i\geq 2$;}\\
(-1)^m, & \textrm{otherwise,}\end{array}\right.
\end{equation}
and $n$ has the canonical factorization $n=p_1^{e_1}p_2^{e_2}\ldots p_m^{e_m}$.
The degree of $\Phi_N(x)$ is given by $\phi(N)$, where $\phi(.)$ is the Euler totient function  \cite{Burton}.

In this paper, a new algorithm for computing a single DFT component, which is based on cyclotomic polynomials, is introduced in Section II. This algorithm is then combined with the Goertzel algorithm in Section III, to produce the most efficient algorithm, in terms of multiplicative complexity (as far as we know), for computing a single DFT component. In Section IV a hardware implementation for the discussed algorithms is proposed and the conclusions are presented in Section V.

\section{The JCO Algorithm}
The computation of a single DFT component, $V_k$, using the JCO algorithm, considers the cyclotomic polynomial $\Phi_L(x)$, where
\begin{equation}
L = \textrm{ord}(W_N^k) = \frac{N}{\gcd(N,k)}.
\end{equation}
Then, by definition, $\Phi_L(x)$ has a zero in $W_N^k$ and $v(x)$ can be written as
\begin{equation} \label{eq:vx_R}
v(x) = \Phi_L(x)Q(x)+R(x),
\end{equation}
where $R(x)$ can be computed by an autoregressive filter and
\begin{equation}
V_k = R(W_N^k).
\end{equation}

Unlike the $p_k(x)$ polynomial, $\Phi_L(x)$ has integer coefficients which, for $L$ smaller than 105, are equal to $0$, $1$ and $-1$\cite{BlahutFA}. Therefore no multiplication is required to compute the polynomial division. The polynomial $\Phi_L(x)$ has degree $\phi(L)$, so that $2[\phi(L)-1]$ real multiplications are needed to compute $V_k$ using the JCO algorithm.

\section{The JCO-Goertzel Algorithm}
The computation of $R(x)$ from $v(x)$ via the JCO algorithm is multiplication free. The polynomial $R(x)$ in \eqref{eq:vx_R} has degree $\leq(\phi(L)-1)$ and can be written as
\begin{equation}
R(x) = p_k(x)q(x)+r(x),
\end{equation}
from which the $V_k$ component can be computed by the Goertzel algorithm, as in  \eqref{eq:Vk_r}.
Therefore, the number of real multiplications in the JCO-Goertzel algorithm is $\phi(L)$. Due to the fact that 
\begin{equation}\label{eq:fi}
\phi(L)< L\leq N,
\end{equation}
it is clear that the JCO-Goertzel algorithm is more efficient, in terms of multiplicative complexity, than the Goertzel algorithm. 


Table \ref{tab:complexity} shows the multiplicative complexity (real multiplications) of the Goertzel, JCO and JCO-Goertzel algorithms, for some values of $N$ and $k$, assuming that $v_n \in \mathbb{R}$. When $\phi(L) = 2$, the cyclotomic polynomial $\Phi_L(x)$ is equal to $p_k(x)$ and the multiplication by the coefficient $A = 2 \cos(2\pi k/N)$ is a trivial one. Consequently, for $L=3,4,6$ in Table \ref{tab:complexity}, the algorithms present the same performance. From \eqref{eq:fi}, it is clear that the only case for which the Goertzel algorithm outperforms  JCO is when $L=N$ and $N$ is a prime number, as indicated in Table \ref{tab:complexity} for $N = 83$.



\begin{table}[hbt]
  \begin{center}
  \caption{Number of Real multiplications required to compute $V_k$, for a length $N$ real sequence $v$, for the Goertzel, JCO and JCO-Goertzel algorithms.}\label{tab:complexity}
  {\scriptsize
  \begin{tabular}{|c|c|c|c|c|c|}
  \hline
  N & k & Goertzel & JCO & JCO-Goertzel & L\\
  \hline
   12 & 1 & 12 & 6 & 4 & 12\\
   & 2 & 2 & 2 & 2 & 6\\
   & 3 & 2 & 2 & 2 & 4\\
   & 4 & 2 & 2 & 2 & 3\\
  \hline
  32 & 1 & 32 & 30 & 16 & 32\\
   & 2 & 32 & 14 & 8 & 16\\
   & 3 & 32 & 30 & 16 & 32\\
   & 4 & 32 & 6 & 4 & 8\\
  \hline
  48 & 1 & 48 & 30 & 16 & 48\\
   & 2 & 48 & 14 & 8 & 24\\
   & 3 & 48 & 14 & 8 & 16\\
   & 4 & 48 & 6 & 4 & 12\\
  \hline
  83 & 1,2,3,4 & 83 & 162 & 82 & 83\\
  \hline
  120 & 1 & 120 & 62 & 32 & 120\\
   & 2 & 120 & 30 & 16 & 60\\
   & 3 & 120 & 30 & 16 & 40\\
   & 4 & 120 & 14 & 8 & 30\\
  \hline
  \end{tabular}
  }
  \end{center}
\end{table}

\section{Hardware Implementation}
The hardware implementation of the Goertzel algorithm can be made using the autoregressive filter 
\begin{equation}
H(z) = \frac{1}{1-W_N^{-k}z^{-1}},
\end{equation}
with input $v_n$, $n = 0,\ldots,N-1$ and output $y_n$ \cite{OppenPDS},\cite{Beraldin_goertzel}. The filter computes
\begin{equation}
V_k = y_N.
\end{equation}

\begin{figure}[hbt]
 \centering
 \includegraphics[scale=0.52]{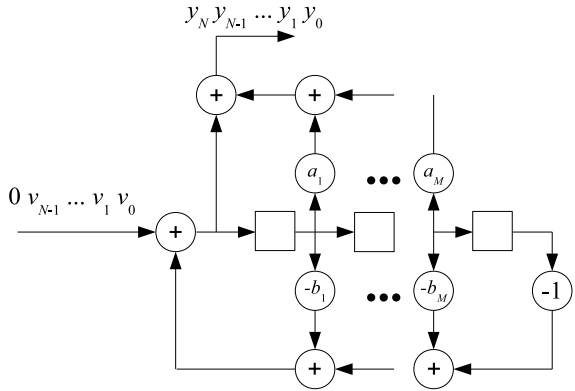}
 \caption{Autoregressive filter to implement the JCO algorithm. The components $v_n$ are fed into arrival order and $M = \phi(L)-1$. The computed DFT component is $V_k = y_N$.}
 \label{fig:SSO}
\end{figure}

To derive a hardware implementation of the JCO algorithm, $H(z)$ is written as
\begin{equation}
H(z) \frac{ \prod_{\textrm{ord}(W_N^i)=L,i\neq k}\left(1-W_N^{-i}z^{-1}\right) } { \prod_{\textrm{ord}(W_N^i)=L,i\neq k}\left(1-W_N^{-i}z^{-1}\right) },
\end{equation}
so that, from \eqref{eq:PHI} and the $\Phi_L(x)$ symmetry,
\begin{equation}
H(z) = \frac{ \prod_{\textrm{ord}(W_N^i)=L,i\neq k}\left(1-W_N^{-i}z^{-1}\right) } { \Phi_L(z^{-1}) }.
\end{equation}
The degrees of the denominator and numerator polynomials are $\phi(L)$ and $\phi(L)-1$, respectively. Therefore, $H(z)$ can be expressed in the form
\begin{equation}
H(z) = \frac{1+a_1 z^{-1}+\ldots +a_{\phi(L)-1}z^{-\phi(L)+1}}{1+b_1 z^{-1}+\ldots +z^{-\phi(L)}}.
\end{equation}
Figure \ref{fig:SSO} shows the general hardware implementation of the JCO algorithm. The multiplications by $a_j$ need to be made once only and the multiplications by $b_j$ are all trivial. An attractive aspect of this implementation is that the $v_n$ components are fed into the shift register circuit in arrival order, thus requiring no components storage.

\begin{figure}[hbt]
 \centering
 \includegraphics[scale=0.45]{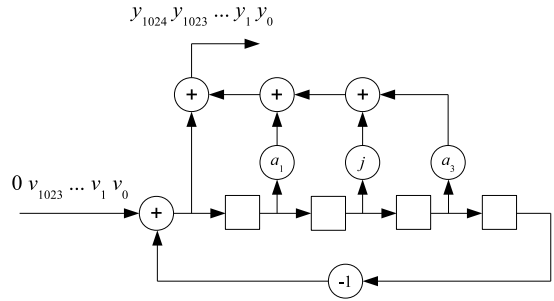}
 \caption{Example 1 JCO hardware implementation, $a_1 = \frac{\sqrt{2}}{2}(1+j)$ and $a_3 = \frac{\sqrt{2}}{2}(-1+j)=-a_1^*$.}
 \label{fig:X128}
\end{figure}

\textbf{Example:} To compute the component $V_{128}$ of a 1024-point DFT, it is necessary to determine the order of $W_{1024}^{128}$. Since $W_{1024}^{128} = W_8$, it has order eight. There are $\phi(8) = 4$ elements of order eight, namely, $W_8$, $W_8^3$, $W_8^5$ and $W_8^7$. Therefore
\begin{equation*}
H(z) = \frac{ (1-W_8 z^{-1})(1-W_8^3 z^{-1})(1-W_8^5 z^{-1}) } { \Phi_8(z^{-1}) }
\end{equation*}
which leads to
\begin{equation*}
H(z) = \frac{ 1+\frac{\sqrt{2}}{2}(1+j)z^{-1}+j z^{-2}-\frac{\sqrt{2}}{2}(1-j)z^{-3} } { 1+z^{-4} }.
\end{equation*}
The desired DFT component is obtained from the filter output as $V_{128} = y_{1024}$. The corresponding hardware implementation of the JCO algorithm is shown in Figure \ref{fig:X128}. The computation of  $V_{128}$ requires only $2$ multiplications and $1027$ additions, in contrast to $1024$ multiplications and $2049$ additions as required by the Goertzel algorithm.

\section{Conclusions}
In this paper two new algorithms for computing a single component of the discrete Fourier transform, the JCO and the JCO-Goertzel algorithms, are proposed. Both algorithms have, in general, a better performance in terms of computational complexity, when compared to the well known Goertzel algorithm, which is the standard procedure for this type of computation. In particular, the JCO-Goertzel algorithm has the lowest multiplicative complexity, as far as we know, of the algorithms that compute a single component of an $N$-point DFT. 

The approach presented in this paper represents a change in paradigm with respect to the Goertzel method in the sense that, instead of using a fixed polynomial of degree 2, the cyclotomic polynomial $\Phi_L(x)$ is used. $L$ is an integer that is a function of $N$, the DFT length, and $k$, the index of the DFT component to be computed. This means that different components of the DFT will be computed with different complexities. Consequently, considering that the JCO-Goertzel algorithm requires less than $N$ multiplications for each DFT component computed, it can be used to compute an $N$-point DFT with less than $N^2$ multiplications. 

Application scenarios that use the Goertzel algorithm will benefit from the techniques introduced in this paper \cite{Chicharo_Goertzel}, \cite{Garcia_sliding_goertzel}. In the field of real numbers, for instance, the detection of DTMF signals is a typical and important application \cite{Beck_Goertzel}. In the finite field context, the syndrome computation in the decoding of a BCH code can be implemented by the Goertzel algorithm \cite{BlahutCOD}. Therefore, considering that a finite field version of the results presented here can be derived following essentially the same approach, the proposed algorithms can be used to assist the decoding of such codes.


\end{document}